\documentclass{aa}  

\usepackage{graphicx}
\usepackage{txfonts}
\usepackage[colorlinks=true,linkcolor=blue,citecolor=blue]{hyperref}
\usepackage{float}
\usepackage{placeins}

\begin{document}

   \title{Magnetic fields in extreme primordial halos: turbulent collapse and implications for early quasar formation}

%   \subtitle{*******}

   \author{V. B. Díaz\inst{1}%\fnmsep\thanks{vdiaz@hs.uni-hamburg.de}
          \and
          D. R. G. Schleicher\inst{2}
          \and
          M. A. Latif\inst{3}%\fnmsep\thanks{latifne@gmail.com}
          %\and
          %P. Grete\inst{1}
          \and
          R. Banerjee\inst{1}
          %\inst{2}\fnmsep\thanks{Just to show the usage
          %of the elements in the author field}
          }

   \institute{Hamburger Sternwarte, Universität Hamburg, Gojenbergsweg 112, D-21029 Hamburg, Germany \\ \email{vanesa.belen.diaz.diaz@uni-hamburg.de}
         \and
             Dipartimento di Fisica, Sapienza Universit\`a di Roma, Piazzale Aldo Moro 5, 00185 Rome, Italy \\ \email{dominik.schleicher@uniroma1.it}
         \and
             Physics Department, College of Science, United Arab Emirates University, PO Box 15551, Al-Ain, UAE \\ \email{latifne@gmail.com}
             }

   \date{Received -; accepted -}

% \abstract{}{}{}{}{} 
% 5 {} token are mandatory
 
  \abstract
   {It is sometimes suggested that the most massive quasars at high redshift may have formed from rare high-sigma peaks in the cosmic density field. We explore here the evolution in a massive primordial halo corresponding to one of these rare sigma peaks, employing cosmological high-resolution magneto-hydrodynamical zoom-in simulations with initial field strength of $10^{-14}-10^{-8}$~G. The dark matter halo forms at the intersection of strongly convergent flows, leading to the formation of highly supersonic turbulence already on intergalactic scales, with turbulent Mach numbers of order $10-20$ also within the halo. Turbulent magnetic field amplification has never been explored in this regime and we therefore investigate whether this gives rise to effects similar to those observed in more typical halos. While the weaker initial field is somewhat more strongly amplified as a result of the shear flows, overall the evolution in the different simulations is rather similar and the flows are dominated by supersonic turbulence. We show in particular that the turbulent Jeans mass always dominates over the thermal and magnetic Jeans masses, and only on scales of $10^{-2}-10$~pc, the magnetic Jeans mass may become comparable to the thermal one. Our simulations thus strongly suggest the evolution to be dominated by the large-scale dynamics. As established in previous work, we thus expect the formation of central massive objects of a few times $10^4$~M$_\odot$ also in the presence of magnetic fields. The situation is somewhat different from more typical atomic cooling halos, where previous results have indicated a larger relevance of the magnetic Jeans mass on intermediate scales, potentially enhancing the mass of the massive object. 
   }


   \maketitle
%
%-------------------------------------------------------------------

\section{Introduction}
The James Webb Space Telescope (JWST) has revealed that supermassive black holes (SMBHs) are present already at early cosmic times \citep{Adamo2025, Maiolino2024, Harikane2023}. They further have shown that the masses of their black holes correspond to high fractions of the stellar mass of their host galaxies, where mass fractions of order $0.1$ have been inferred \citep[e.g.][]{Maiolino2024}, corresponding to a clear enhancement with respect to the relation in the local Universe \citep{Kormendy2013}. While it was already a challenge to explain the black hole masses of previous quasars detected at $z\sim6-7$ \citep[e.g.][]{banados, morlock}, the problem became essentially more serious with recent discoveries. 

Several different formation channels have been proposed for high redshift black holes, following the ideas laid out by \citet{Rees1984}. The possible scenarios include direct collapse \citep{Koushiappas2004, brom, begelman, Wise2008, Schleicher2013, Ferrara2014}, runaway collisions in dense stellar clusters \citep{PortegiesZwart2007, devicchi1, Katz2015, Reinoso2018, Escala2021, Vergara2023, Vergara2024, Rantala2024}, mergers in black hole clusters in the presence of gas inflows \citep{Lupi2014, Kroupa2020, Gaete2024} as well as mixed scenarios involved accretion and collisions \citep{tjarda, Tagawa2020, Chon2025, Schleicher2023, Reinoso2023, Saavedra2024,  Solar2025}.

We focus here particularly on the possibility of direct collapse, which makes it possible to form particularly massive seeds in massive primordial halos \citep{brom, latif2,Latif2013, latif2016, Schober2012, Schober2013}. The dynamics in these halos often involve highly supersonic turbulence formed as a result of virialisation \citep{Wise2007, Greif2008}, which has been shown to efficiently amplify the magnetic fields in these halos \citep{Schleicher2010, Sur2010, Sur2012, Turk2012, latif7, latif8}. Such magnetic fields can potentially also affect the dynamics, leading to angular momentum transport on large scales due to magnetic braking as well as the suppression of fragmentation due to magnetic pressure \citep{sharda1,sharda2, Latif2022, Latif2023} or the presence of a magneto-rotational instability on the disc scale \citep{Silk2006}. The feeding of supermassive black holes can be significantly enhanced in the presence of strong magnetic fields \citep{Begelman2023}.

However, the dynamical role of magnetic fields is not uniform across all environments. In Population~III (Pop~III) star-forming minihalos, %where turbulence is tipically subsonic to transonic, 
magnetic fields can be amplified during gravitational collapse \citep{Mackee2020,Sadanari2023,Higashi2024}. While several studies find that fields amplified to near-equipartition suppress fragmentation and favour the formation of more massive stars \citep{sharda1,sharda2,Stacy2022,Saad2022,Sadanari2021,Sadanari2024}, or drive the merging of fragments into a single massive protostar through magnetic torque \citep{Machida2025}, others find that magnetic pressure never exceeds thermal pressure and that magnetic fields are therefore largely unimportant for the fragmentation of primordial gas \citep{Prole2022}. More recent radiation-magnetohydrodynamics simulations find that magnetic fields instead limit the maximum stellar mass through reduced compressional heating and suppressed mass transport \citep{Sharda2025a,Sharda2025b,Vanveenen2026}. The origin of these discrepancies remains debated, suggesting that the dynamical role of magnetic fields depends sensitively on the specific physical conditions of each environment as well as the length and density scales under consideration. 

The magnetic field amplification in massive primordial atomic cooling halos has already been looked at in various previous studies, including \citet{latif7, grete2019, Hirano2021, Hirano2022, Latif2022, diaz}. Even in the presence of supernova-driven turbulence, a small-scale dynamo was shown to be operational \citep{Seifried2014}. Here, we focus on a particularly extreme situation previously explored by \citet{LatifNature}, focusing on a particularly high sigma-peak in a cosmological simulation, given rise to a dark matter (DM) halo that starts out with $3.0\times10^5$~M$_\odot$ at $z=35$, which grows up to $1.4\times10^{12}$~M$_\odot$ at $z=6$. This halo forms at the intersection of cosmic filaments, given rise to very efficient mass flows towards the halo as well as driving highly supersonic turbulent flows even on scales of the intergalactic medium in the environment of the halo. The strong supersonic turbulence in this environment initially prevents gravitational collapse, until sufficient gas accumulates to lead to a sudden burst of collapse, producing infall rates of $\sim0.55$~M$_\odot$~yr$^{-1}$, thus providing ideal conditions for the formation of a central very massive object.

Here, we explore the evolution of magnetic fields in an extreme environment like this. Given the strongly supersonic nature of the flows, it will be difficult to assess the presence of a turbulent dynamo, which is known to require very high resolution \citep[e.g.][]{Sur2010, FederrathSur2011, Turk2012}, but rather focus on the implications of a magnetic field generated through an astrophysical mechanism to assess its potential implications in the large-scale dynamics of the halo. Whether magnetic fields play a comparably relevant dynamical role in this extreme halo, same as shown in more typical atomic cooling halos \citep[e.g.][]{Latif2022}, is not trivial. One could on the one hand assume that this would be the case, as there is even stronger turbulence that could more efficiently amplify the magnetic field through the small-scale dynamo. On the other hand, highly supersonic regimes are known to disfavour the efficiency of the magnetic field amplification via this dynamo process \citep{Federrath2011,Schober2015}. We investigate this question through the simulations presented in this work.

The structure of this paper is as follows: In section~\ref{methods}, we present our methodology. In section~\ref{results}, we provide the results from our simulations. A summary and discussion is provided in section~\ref{summary}.

%--------------------------------------------------------------------
\section{Computational methods}\label{methods}

The simulations presented here mostly follow the methodology presented by \citet{diaz} applied to the specific computational setup developed by \citet{LatifNature}. The cosmological magneto-hydrodynamics (MHD) zoom-in simulations are conducted using the 3D MPI-parallel, Eulerian, block structured, adaptive mesh refinement (AMR) code ENZO \citep{Bryan2014,enzo2}, an open-source code including self-gravity as well as a wide range of possible sub-grid and microphysical processes. It solves the continuity, Euler, energy and induction equations  of cosmological (comoving) ideal MHD as well as the Poisson equation for self-gravity. The Riemann problem is solved using the Harten-Lax-van Leer (HLL) Riemann solver \citep{Toro1997}, while the code takes care of the divergence constraint $\nabla\cdot B=0$ using a wave-like hyperbolic cleaner \citep{Dedner2002}. The reconstruction of  the variables is pursued using the piecewise linear method (PLM) \citep{vanleer1979}. For the purpose of comparison with other studies, we briefly mention that the Riemann solver employed here  is somewhat more diffusive compared to the HLL3R scheme used in similar works with the Flash code \citep[see e.g.][]{Sur2010,FederrathSur2011,Sur2012}; which effectively decreases the magnetic Reynolds number of the flow and makes it harder to capture potential turbulent amplification effects \citep{Turk2012,latif7}.

We note that the solver employed here differs from that used by \citet{LatifNature}, who used the piecewise parabolic method (PPM) with the HLLC Riemann solver. PPM uses a third-order accurate parabolic interpolation scheme, whereas PLM uses a second-order linear interpolation, making PLM somewhat more numerically diffusive. Similarly, the HLL solver does not resolve the contact wave, making it more diffusive than HLLC. These differences are an expected and necessary consequence of including MHD in ENZO. The increased numerical diffusivity may result in smoother discontinuities compared to \citet{LatifNature}, but does not affect the global collapse dynamics.

To generate cosmological initial conditions at $z=150$ we use MUSIC \citep{Hahn2011} using the second-year Planck cosmological model using lowP + lensing + BAO + JLA + H$_0$ cosmological parameters \citep{planck2016} in a cosmological box of $25$~Mpc~h$^{-1}$. As there are only a few dozen halos per Gpc$^3$ corresponding to the rare sigma peaks that are being considered here, \citet{LatifNature} simulated many such boxes using different random seeds until they found one that would form a halo with $1.4\times10^{12}$~M$_\odot$ at $z=6$ primarily by accretion from cosmic filaments rather than major mergers. The computational box in which this halo was found is the one employed in the present work, and we use the Rockstar halo finder \citep{rockstar} to extract its properties listed in Table~\ref{sims}. The setup includes a top-grid resolution of $256^3$ cells with five additional nested refined levels within the Lagrangian volume of the halo, yielding a maximum DM resolution of $3636$~M$_\odot$, an effective initial spatial resolution of $8192^3$ cells, and $975$ million DM particles. DM particles are smoothed at refinement level 12 to avoid spurious heating of the baryons. We further allow for 25 dynamical levels of refinement in the innermost nested grid, resulting in a maximum spatial resolution of approximately $20$~AU at the highest refinement level. Our refinement criteria are based on baryonic overdensity with a super-Lagrangian refinement exponent of $-0.2$, the Jeans length and the DM mass as criteria for further refinement \citep[see also][]{LatifNature}. We employ a Jeans resolution of 64 cells per Jeans length, which is known to be sufficient to resolve turbulent eddies in cosmological simulations \citep{FederrathSur2011,Latif2013b} but may not fully capture small-scale dynamo amplification \citep{diaz}.

When the maximum refinement level is reached, we employ a pressure floor on the highest refinement level to avoid artificial fragmentation and to make the cells Jeans stable \citep{Machacek}, satisfying the Truelove criterion \citep{Truelove1997}. Specifically, this is achieved by defining, for each cell, an artificial pressure which is the maximum of the thermal pressure and $KG\rho_b^2\Delta x^2/\mu$, where $\rho_b$ is the local baryon density, $\Delta x$ is the cell width and $\mu$ is the mean particle mass. We additionally set the dimensionless constant $K$ to 100, ensuring a Jeans length to cell size ratio of 10 \citep{LatifNature,Latif2022}. This approach allows us to follow the evolution of the halo beyond the onset of collapse.

With this setup, we perform three simulations with initial uniform magnetic field strengths of $B_0=10^{-14}$, $10^{-10}$ and $10^{-8}$~G (proper), all of which are consistent with the upper limits for primordial magnetic fields established by the Planck mission \citep{Planckmagnetic}. This choice is motivated by the fact that starting from a weak field seed alone makes it difficult to achieve the high Reynolds numbers required in numerical simulations, potentially leading to an underestimate of the growth of the magnetic field and its potential dynamical impact \citep{Schober2012PhysRev}. We select three different evolutionary stages corresponding to peak densities of $\rho_{peak}=3\times10^{-15}$ , $10^{-13}$ and $10^{-12}$~g~cm$^{-3}$, representing an evolution of approximately $5-13$~kyr beyond the initial collapse, depending on the simulation.

\subsection{Chemical model}
Here we use the standard-treatment of primordial chemistry in ENZO following the same setup as \citet{LatifNature}, with no Lyman-Werner (LW) sources present. We solve the rate equations of nine different chemical species ($\mathrm e^-$, $\mathrm H^-$, H, $\mathrm H^+$, He, He$^+$, He$^{++}$, $\mathrm H_2$, $\mathrm H_2^+$) \citep{Anninos1997, Abel1997}. Our model includes H$_2$ cooling rates valid at both low and high densities, where local thermodynamic equilibrium effects become important \citep{Glover2008}. These rates further include the transition from optically thin to optically thick cooling as H$_2$ becomes opaque to its own emission lines at higher densities \citep{Ripamonti2004}. Our model also includes cooling from collisional ionization and excitation of H and He, bremsstrahlung radiation, inverse Compton scattering by the cosmic microwave background, and radiative recombination.

Following \citet{LatifNature}, we do not include collisional ionizations of H by H and contributions to the H$_2$ collisional dissociation rate by dissociative tunnelling, as these processes become relevant primarily in the presence of strong LW radiation fields \citep{glover1,glover2}, which are absent in our simulations. While \citet{LatifNature} justify this omission by noting that temperatures in their halo remain below the relevant thresholds ($\sim8000$~K and $\sim2000-4500$~K for the two reactions, respectively), we note that our simulations reach temperatures above $2000$~K at large scales, with a peak of $\sim7000$~K at $\sim300$~pc. In this region, the dissociative tunnelling rate may therefore become non-negligible. However, given that this occurs far from the collapsing core, in low-density material where H$_2$ has not yet formed efficiently, we expect the impact on our main results to be limited.

%-----------------------------------------------------------------
\section{Results}\label{results}

In this section we present the results of our cosmological MHD simulations of the halo identified in \citet{LatifNature}. At a redshift of $z\sim25.6$, the halo has a DM mass of about $5\times10^7$~M$_\odot$. We performed three simulations with different initial magnetic field strengths of $10^{-14}$~G, $10^{-10}$~G, and $10^{-8}$~G (proper) using a Jeans resolution of 64 cells per Jeans length.

Following the methodology of \citet{LatifNature}, we imposed a pressure floor at 25 refinement levels (the highest refinement level of our simulations) to prevent the gas from collapsing further, ensuring the Truelove criterion for the Jeans length \citep{Truelove1997}. We then select three different evolutionary stages corresponding to peak densities of $3\times10^{-15}$~g~cm$^{-3}$, $10^{-13}$~g~cm$^{-3}$ and $10^{-12}$~g~cm$^{-3}$, which represent an evolution of approximately $5-13$~kyr depending on the simulation, to follow the halo collapse.

 We begin by analysing the physical properties of the collapsing halo in Sect.~\ref{physical} and then, in Sect.~\ref{magnetic}, we focus on the magnetic properties to determine whether magnetic fields affect the dynamics of the gas in this particular scenario.

\begin{table}%[H]
\caption{Properties of the simulated halos shortly after the onset of collapse, when reaching a peak density of $\sim 3\times10^{-15}$~g~cm$^{-3}$. Listed are the initial magnetic field seed strength $B_0$ in proper units, the gas mass $M_\mathrm{gas}$, the total halo mass $M_\mathrm{total}$ enclosed within the virial radius $R_\mathrm{vir}\approx 0.455$~kpc, the thermal Jeans mass $M_{J,th}$ evaluated at $R_\mathrm{vir}$, the normalized mass-to-flux ratio $\mu$, the spin parameter, and the redshift $z$.} % title of Table
\setlength{\tabcolsep}{3.2pt}
\resizebox{\columnwidth}{!}{
\label{tab:table4}      % is used to refer this table in the text
\centering    % used for centering table
\begin{tabular}{ c c c c c c c %@{\hspace{1cm}}
}        % centered columns (4 columns)
\hline\hline                 % inserts double horizontal lines
$B_0$ & $M_\mathrm{gas}$ & $M_\mathrm{total}$&
$M_\mathrm{J,th}$ &
$\mu$ & 
Spin & 
$z$ \\

[G] & 
[$\mathrm{M}_\odot$] & [$\mathrm{M}_\odot$] &[$\mathrm{M}_\odot$]\\
% table heading 
\hline                        % inserts single horizontal line
$10^{-14}$  & 
$5.60\times10^{6}$ & 
$4.82\times10^{7}$ & 
$8.06\times10^{6}$ & 
163.08&
0.024 &  
25.623\\
$10^{-10}$  & 
$5.65\times10^{6}$ & 
$4.83\times10^{7}$ & 
$6.95\times10^{6}$ & 
14.97&
0.024 &  
25.616\\
$10^{-8}$  & 
$5.64\times10^{6}$ & 
$4.83\times10^{7}$ & 
$6.95\times10^{6}$ & 
9.23&
0.024&  
25.615\\
\hline                                   %inserts single line
\end{tabular}
}\label{sims}
\end{table}

%-------------------------------------------------------------
%  Amplification and Saturation of magnetic fields (subsec 1)
%-------------------------------------------------------------

%\subsection{Amplification and saturation of the magnetic field}\label{high}

%-------------------------------------- Two column figure (place early!)
   \begin{figure*}
   \centering
   \includegraphics[width=\textwidth]{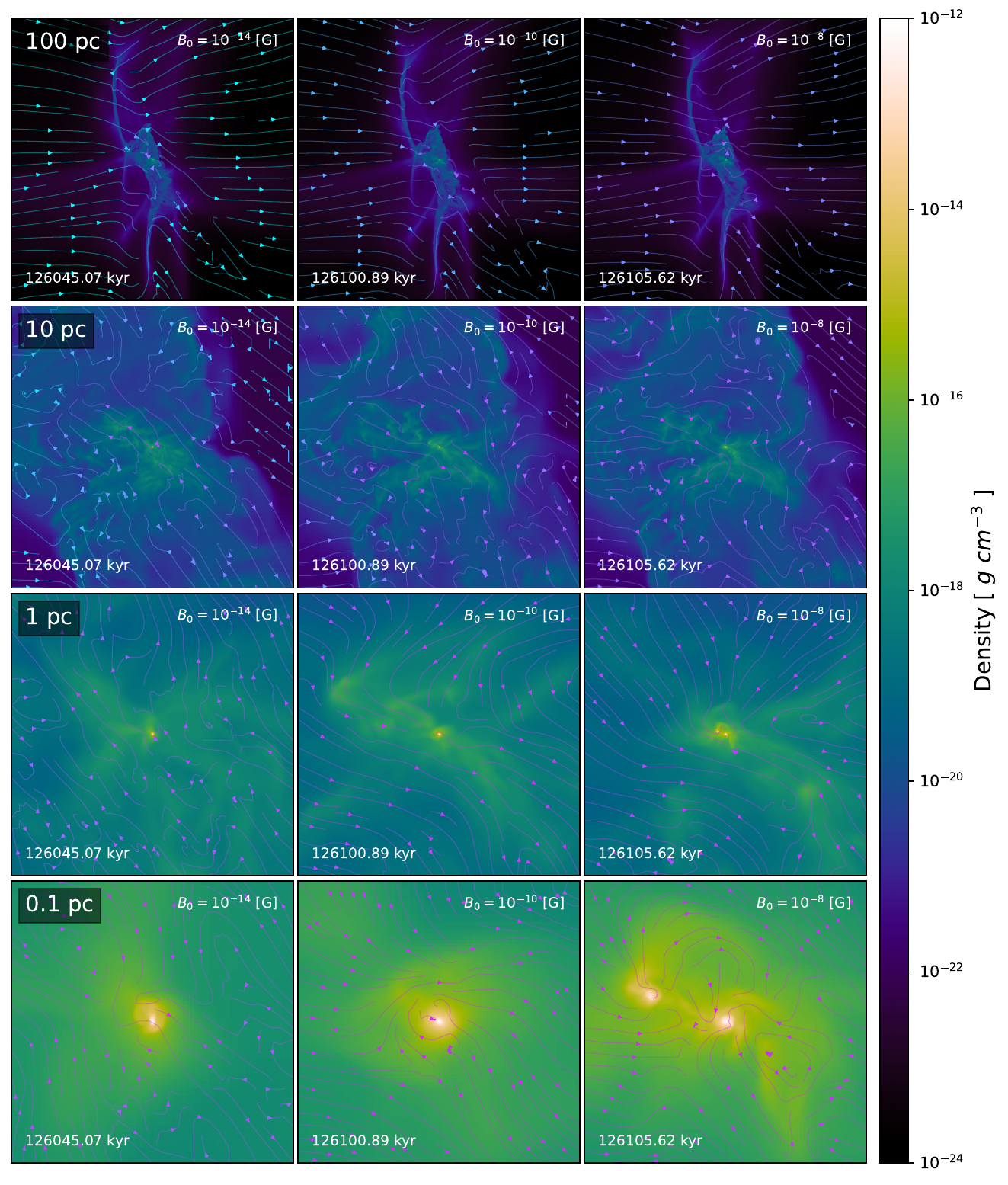}
    \caption{Density-weighted projections of the density along the y-axis at four different spatial scales. From top to bottom the physical width was varied using 100, 10, 1 and 0.1~pc, and from left to right we varied the initial magnetic field strength to $B_0=10^{-14}$, $10^{-10}$ and $10^{-8}$~[G] (proper). All projections correspond to the final evolutionary stage i.e. when the halo reaches a peak density of $1\times10^{-12}$ $\mathrm{[g\,cm^{-3}]}$. Times are indicated in the bottom left of each panel, and magnetic field lines are colour-coded from cyan to magenta with increasing field strength. The overall density morphology and the characteristic four cold stream pattern remain mostly unaffected by $B_0$ on large scales. Differences only appear at smaller scales, where the strongest field case develops two distinct cores.}
    \label{fig:densityproj}
    \end{figure*}
%-----------------------------------------------------------------

%-------------------------------------- Two column figure (place early!)
   \begin{figure*}
   \centering
   \includegraphics[width=\textwidth]{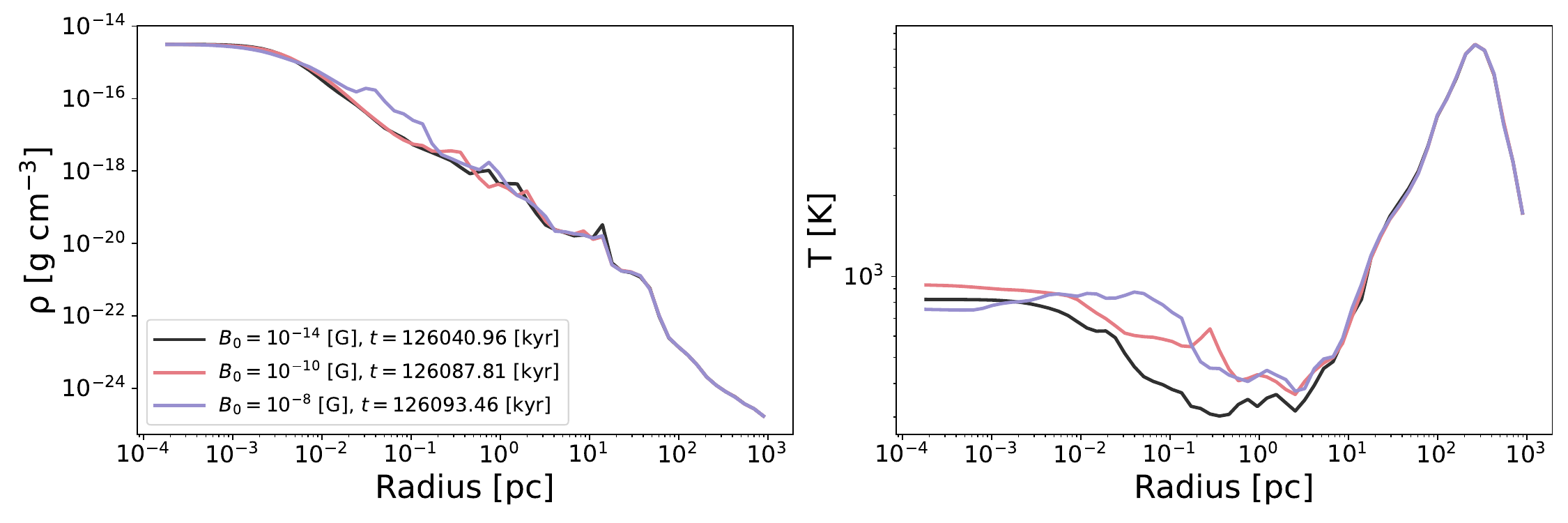}
   \caption{Mass-weighted spherically binned radial profiles of density and temperature %, radial velocity, turbulent velocity, sonic Mach number, turbulent Mach number, mass accretion rate for the halo 
   when reaching a peak density of $\rho_{peak}=3\times10^{-15}$~$\mathrm{[g\,cm^{-3}]}$. Black, pink and purple lines are for runs with $B_0=10^{-14}$, $10^{-10}$ and $10^{-8}$~[G] (proper), respectively. The density profile closely follows an isothermal collapse profile, while the temperature profile show the expected H$_2$ dominated cooling below $10$~pc. This occurs for all three initial magnetic field strengths.}
    \label{fig:physprop64-1}
    \end{figure*}
%-----------------------------------------------------------------
%-------------------------------------- Two column figure (place early!)
   \begin{figure*}
   \centering
   \includegraphics[width=\textwidth]{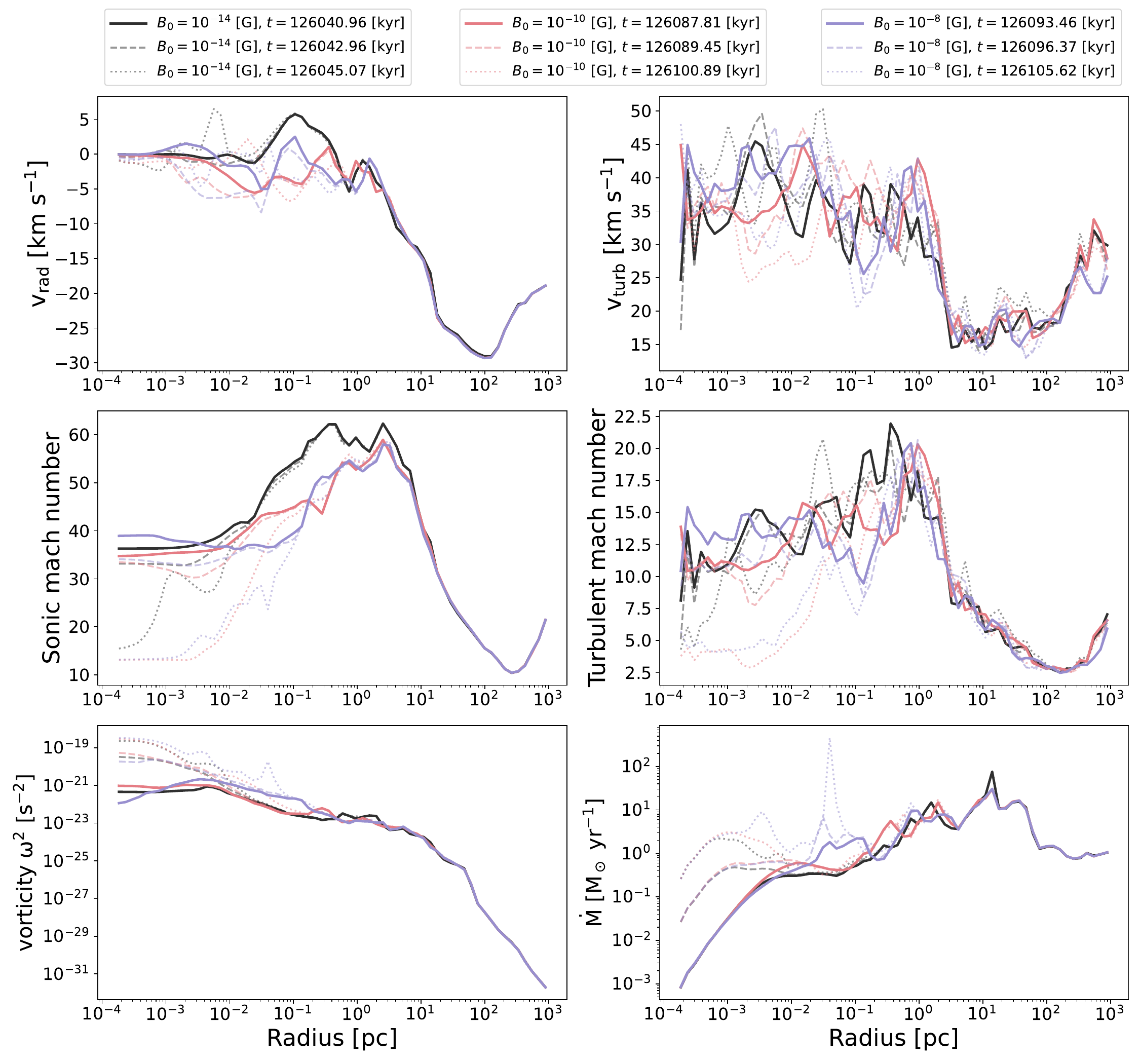}
   \caption{Mass-weighted spherically binned radial profiles of %density, temperature, 
   radial velocity, turbulent velocity, sonic Mach number, turbulent Mach number, mass accretion rate for the halo when reaching three different peak densities. Black lines are for runs with $B_0=10^{-14}$~[G] (proper), pink lines for runs with $B_0=10^{-10}$~[G] (proper), and purple lines for runs with $B_0=10^{-8}$~[G] (proper). Solid lines correspond to $\rho_{peak}=3\times10^{-15}$~$\mathrm{[g\,cm^{-3}]}$, dashed lines correspond to $\rho_{peak}=1\times10^{-13}$~$\mathrm{[g\,cm^{-3}]}$ and dotted lines correspond to $\rho_{peak}=1\times10^{-12}$~$\mathrm{[g\,cm^{-3}]}$. The physical properties of the halo show no significant dependence on $B_0$, demonstrating that the collapse dynamics are dominated by the highly supersonic turbulence ($\mathcal{M}\sim 30-60$) driven by the converging cold streams.}
    \label{fig:physprop64}
    \end{figure*}
%-----------------------------------------------------------------

%-------------------------------------- Two column figure (place early!)
   \begin{figure*}
   \centering
   \includegraphics[width=\textwidth]{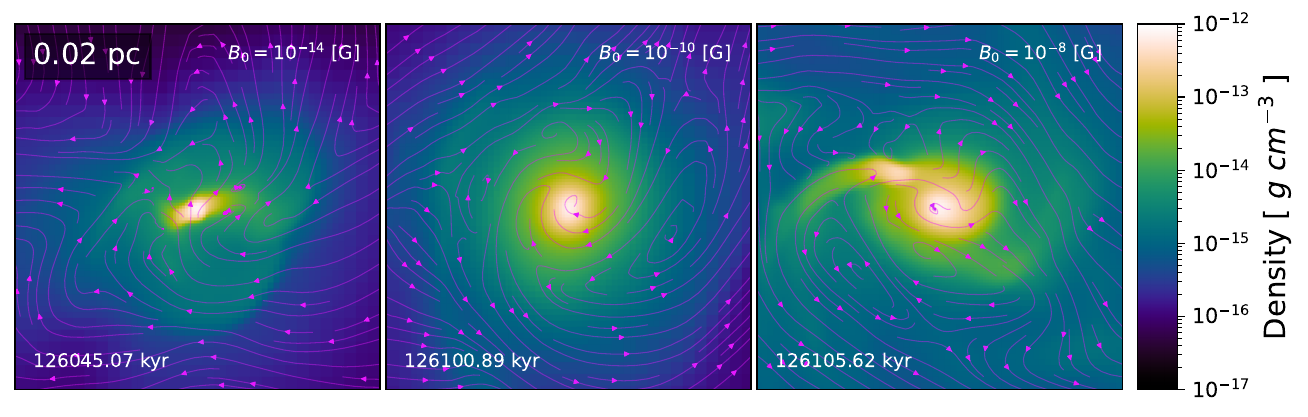}
    \caption{Density-weighted projections of the density along the x-axis with a physical width of $0.02$~pc. From left to right we varied the initial magnetic field strength to $B_0=10^{-14}$, $10^{-10}$ and $10^{-8}$~[G] (proper). All projections correspond to the final evolutionary stage i.e. when the halo reaches a peak density of $1\times10^{-12}$ $\mathrm{[g\,cm^{-3}]}$. Times are indicated in the bottom left of each panel, and magnetic field lines are shown in magenta. Disc-like structures are present at sub-pc scales, with the strongest initial magnetic field case showing the clearest disc-like features. The morphology varies between the three simulations but is consistent with the dynamically subdominant role of the magnetic field, as the disc formation is not suppressed.}
    \label{fig:densityproj-002pc}
    \end{figure*}
%-----------------------------------------------------------------

%-------------------------------------- Two column figure (place early!)
   \begin{figure}
   \centering
   \includegraphics[width=\hsize]{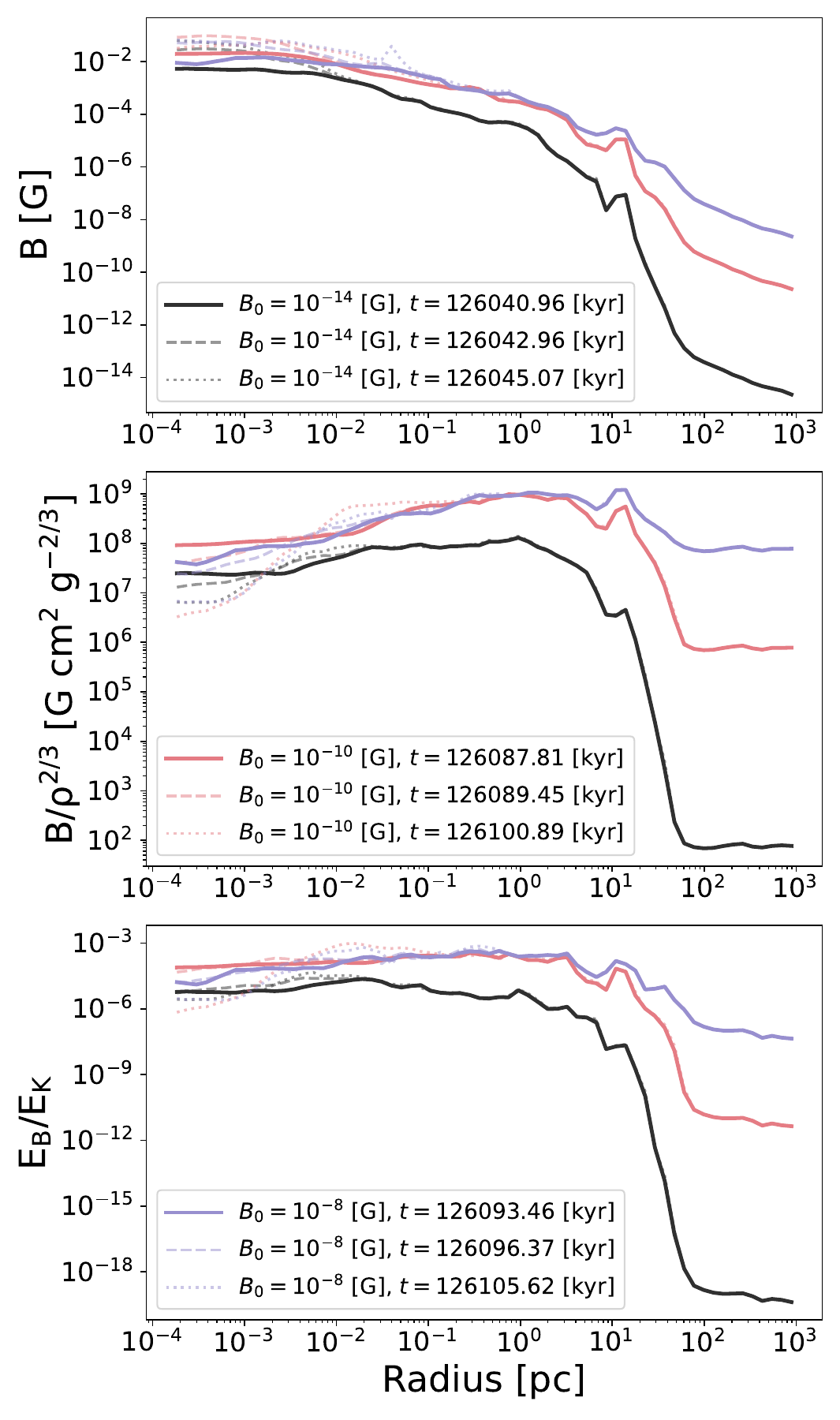}
   \caption{Mass-weighted spherically binned radial profiles of magnetic field strength, magnetic field amplification $B/\rho^{2/3}$ and the magnetic-to-kinetic energy ratio $E_B/E_K$ when reaching three different peak densities. Black lines are for runs with $B_0=10^{-14}$~[G] (proper), pink lines for runs with $B_0=10^{-10}$~[G] (proper), and purple lines for runs with $B_0=10^{-8}$~[G] (proper). Solid lines correspond to $\rho_{peak}=3\times10^{-15}$~$\mathrm{[g\,cm^{-3}]}$, dashed lines correspond to $\rho_{peak}=1\times10^{-13}$~$\mathrm{[g\,cm^{-3}]}$ and dotted lines correspond to $\rho_{peak}=1\times10^{-12}$~$\mathrm{[g\,cm^{-3}]}$. The magnetic field is amplified beyond what is expected from pure spherical compression in all three simulations, though it never becomes dynamically relevant, with $E_B/E_K \ll1$ throughout the halo. All three cases converge to similar central field strengths of $B\sim10^{-2}$~G by the latest evolutionary stage.}
    \label{fig:magnetic64}
    \end{figure}
%-----------------------------------------------------------------

%-------------------------------------- Two column figure (place early!)
   \begin{figure*}
   \centering
   \includegraphics[width=\textwidth]{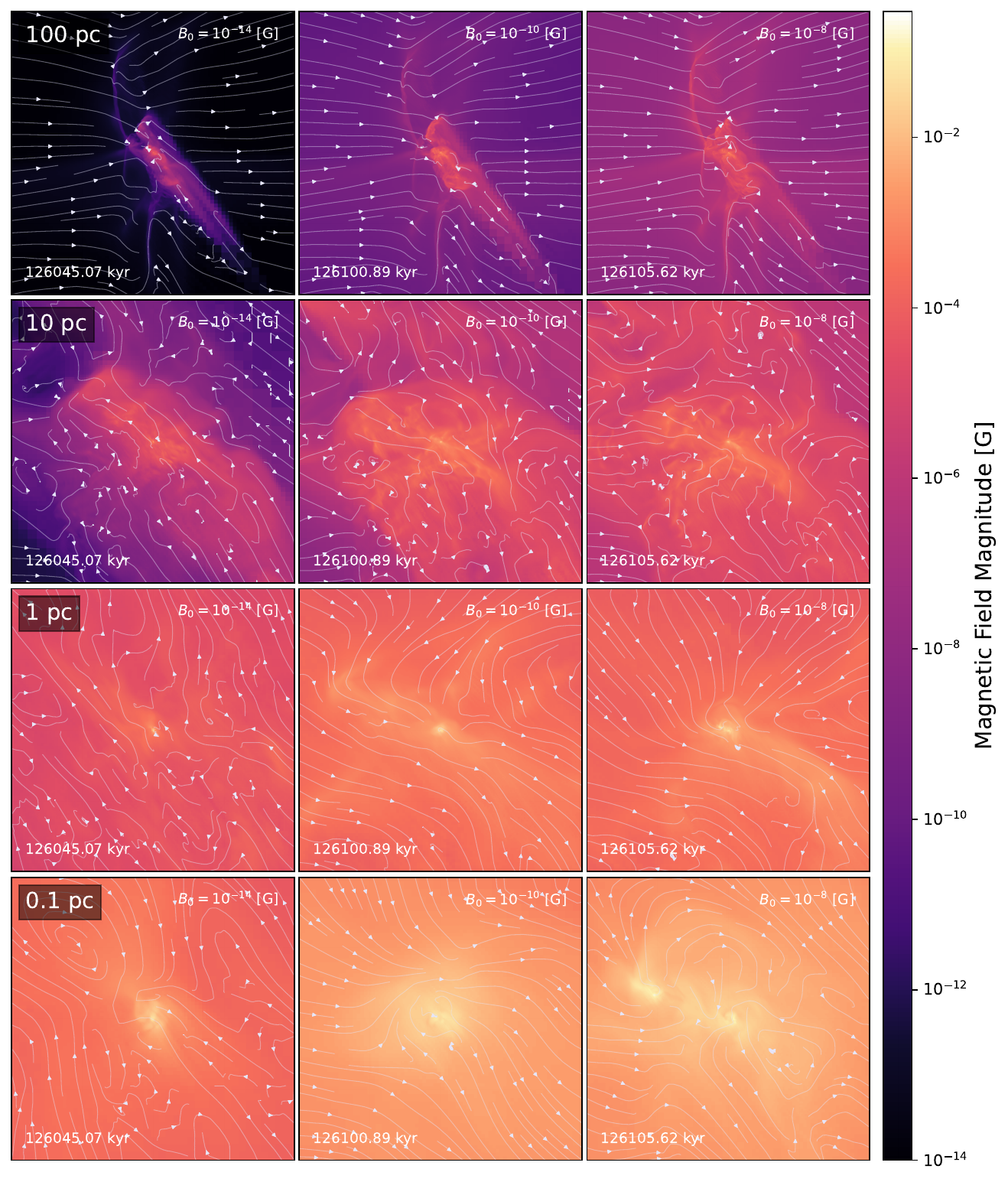}
    \caption{Density-weighted projection of magnetic field magnitude along the y-axis at four different spatial scales. From top to bottom the physical width was varied using 100, 10, 1 and 0.1~pc, and from left to right we varied the initial magnetic field strength to $B_0=10^{-14}$, $10^{-10}$ and $10^{-8}$~[G] (proper). All projections correspond to the final evolutionary stage i.e. when the halo reaches a peak density of $1\times10^{-12}$ $\mathrm{[g\,cm^{-3}]}$. Times are indicated in the bottom left of each panel, and magnetic field lines are shown in white. The magnetic field distribution closely follows the density structure at all spatial scales, with all three cases showing comparable field strengths in the central region. A diagonal structure is visible at $100$~pc which may trace the interface between two of the converging cold flows.}
   \label{fig:bproj}
    \end{figure*}
%-----------------------------------------------------------------

%-------------------------------------- Two column figure (place early!)
   \begin{figure*}
   \centering
   \includegraphics[width=\textwidth]{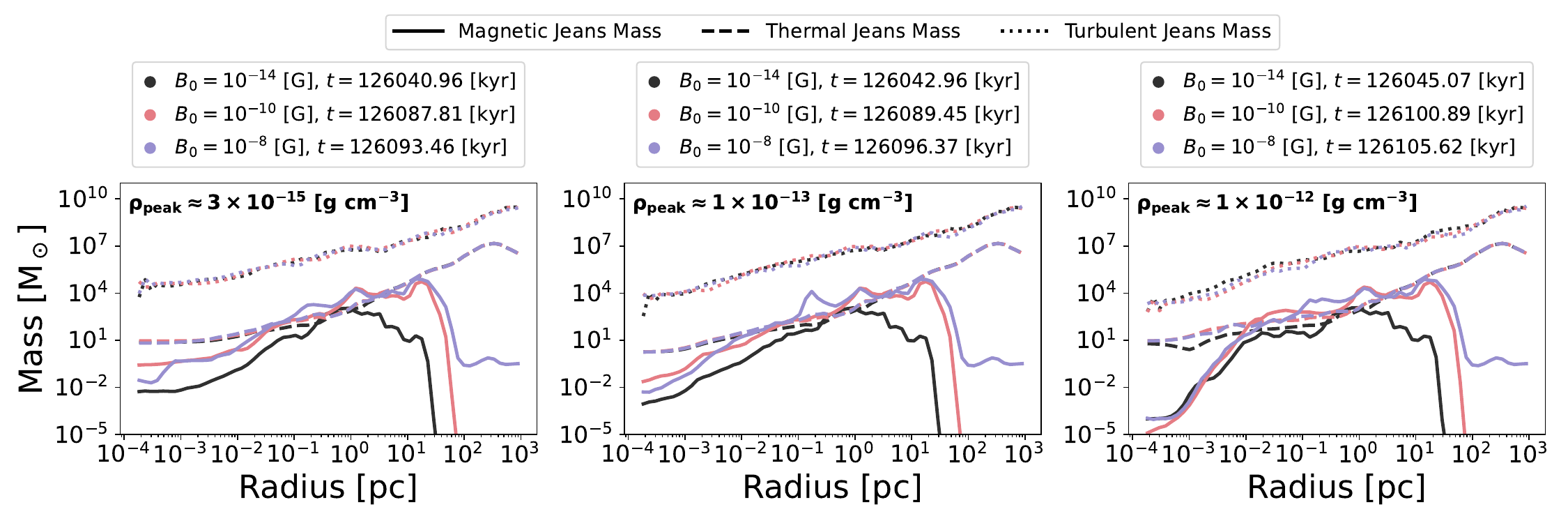}
   \caption{Mass-weighted spherically binned radial profiles of magnetic Jeans mass (solid lines), thermal Jeans mass (dashed lines) and turbulent Jeans mass (dotted lines). Each panel represents different evolutionary stages of the halo that means that the peak density is varied. Colours follow the same convention as previous figures where black lines are for simulations with a proper initial magnetic field strength of $B_0=10^{-14}$~[G], pink lines for simulations with $B_0=10^{-10}$~[G], and purple lines for simulations with $B_0=10^{-8}$~[G]. The turbulent Jeans mass dominates over both thermal and magnetic contributions reaching $\sim10^4$~M$_\odot$ in the central regions. Magnetic pressure provides only a minor contribution in a limited radial range, confirming that the characteristic mass scale for collapse is set by turbulence rather than by magnetic fields.}
    \label{fig:mass64}
    \end{figure*}

\subsection{Physical properties of the halo}\label{physical}

Table~\ref{sims} summarizes the key properties of the halo shortly after the onset of collapse for each initial magnetic field strength. The virial radius is $R_\mathrm{vir}\approx 0.455$~kpc in all three cases, with a gas mass of $M_\mathrm{gas}\approx 5.6\times10^6$~M$_\odot$ and a total halo mass of $M_\mathrm{total} \approx 4.8\times10^7$~M$_\odot$, where $M_\mathrm{total}$ is the sum of dark matter and gas masses enclosed within $R_\mathrm{vir}$. The thermal Jeans mass \citep{Jeans1928,GalacticDynamics2008} is defined as 
\begin{equation}
    M_\mathrm{J,th} = \frac{\pi}{6}\frac{c_s^3}{G^{3/2}\rho^{1/2}},
    \label{thermaljeans_eq}
\end{equation}
where $c_s$ is the sound speed, $\rho$ is the gas density, and $G$ is the gravitational constant. Evaluated at $R_\mathrm{vir}$, this gives $M_\mathrm{J,th} \approx 7-8\times10^6$~M$_\odot$, slightly above the gas mass alone. We note, however, that dark matter provides additional gravity at this scale, which allows the gas to collapse in spite of $M_\mathrm{gas} < M_\mathrm{J,th}$.
The normalized mass-to-flux ratio $\mu$ \citep{Mckee2007} is defined as
\begin{equation}
    \mu = \frac{M_\mathrm{total}}{\pi R_\mathrm{vir}^2 \langle B\rangle_m}\left(\frac{M}{\Phi}\right)_\mathrm{crit}^{-1},
\end{equation}
where $\langle B \rangle_m$ is the mass-weighted average magnetic field strength within $R_\mathrm{vir}$, $\Phi$ is the magnetic flux, and $(M/\Phi)_\mathrm{crit} = (2\pi\sqrt{G})^{-1}$ \citep{NakanoNakamura1978}. The values of $\mu$ decrease with increasing $B_0$, with $\mu > 1$ in all three cases, indicating that gravity dominates over magnetic support at the virial scale.

Figure~\ref{fig:densityproj} shows the density-weighted projection of the density along the y-axis in the central region of the halo at four different spatial scales when the simulation reaches a peak density of $10^{-12}$~g~cm$^{-3}$. From left to right the initial magnetic field increases and from top to bottom we zoom in into the centre. At 100 pc one can see the four cold flows converging toward the centre from different directions. This structure is visible in all the simulations with the same characteristics as shown by \citet{LatifNature}. Rather than having a spherical symmetry, the halo shows an irregular morphology. At 10 pc, the morphology remains qualitatively similar across all cases, showing the irregular and filamentary structure of the halo. Overall, the large scale distribution of the density is thus not affected by the magnetic field. At the 1 pc scale, differences begin to emerge between the cases, though the overall structure remains qualitatively similar. All three simulations show a central concentration of density with a filamentary structure extending outward. We note that our simulation with the strongest initial magnetic field strength starts to form two cores in the centre instead of one. While magnetic fields may thus potentially influence the small-scale structure,  they do not fundamentally alter the collapse dynamics at this scale. Also at 0.1 pc, while the simulations with $B_0\leq10^{-10}$~G show a single clump in the gas distribution, similar to what \citet{LatifNature} found in the pure hydrodynamical simulation at a comparable evolutionary stage, for $B_0=10^{-8}$~G  two different clumps are visible. This binary structure is not observed in \citet{LatifNature} at this early stage of the collapse, where two massive clumps only appear after approximately $100$~kyr of further evolution. Given that the simulations presented here are computationally expensive and we only have low-number statistics, we currently cannot determine whether this binary structure has a statistical or physical origin. 

Figure~\ref{fig:physprop64-1} provides the mass-weighted radial profile of density and temperature at the earliest evolutionary stage ($\rho_{peak}=3\times10^{-15}$~$\mathrm{g\,cm^{-3}}$) for the three different initial magnetic field strengths. 

The density profile shows a similar behaviour for all three initial magnetic field strengths at each evolutionary stage that approximates an isothermal density profile, with some occasional peaks in case of substructure within the collapsing gas. This behaviour is commonly found as the result of gravitational collapse in such halos and has also been found by \citet{LatifNature} in their purely hydrodynamic simulations of the same halo. At scales smaller than $10^{-2}$~pc, the profiles flatten as the gas accumulates in the central regions. On the scale between $10^{-2}-10^{-1}$~pc, a density enhancement is visible in the $B_0=10^{-8}$~[G] run at this early stage, suggesting the mass is already concentrating in this region, which later develops into a second clump at more advanced evolutionary stages, as shown in Figure~\ref{fig:densityproj}.

The temperature profile shows the thermal behaviour expected for halos where cooling is dominated by molecular hydrogen (H$_2$). After an initial temperature peak of $\sim7000$~K due to the virialization shocks, H$_2$ cooling keeps the temperature below 1000~K from the scale of 10~pc down to the centre. The simulations with the three different initial magnetic fields show rather similar temperature profiles that strongly overlap on scales above $5$~pc, while then the evolution becomes slightly different on smaller scales in the non-linear regime.

Figure~\ref{fig:physprop64} extends the analysis of the physical properties to three evolutionary stages, showing the mass-weighted radial profile of the radial velocity, turbulent velocity, sonic and turbulent Mach numbers, vorticity squared, and mass accretion rate.
The radial velocity shows a clear infall at large scales ($r>1$~pc), reaching a maximum infall velocity of $\sim30$~km~s$^{-1}$ at $100$~pc, consistent with the strong convergent flows shown in Figure~\ref{fig:densityproj}. At intermediate scales ($10^{-3}$~pc~$<r<1$~pc) the infall velocity decreases to values of $5-10$~km~s$^{-1}$, showing oscillations that reflect the irregular and filamentary structure of the gas rather than spherical symmetry. Within the central region it approaches zero as the very central region is not collapsing yet.

In the turbulent velocity, we can clearly see two different velocities if we compare the outer with the inner part of the halo. At a scale of $10^3$~pc the turbulent velocity is approximately 30 km/s, which then starts to decrease towards intermediate scales reaching a minimum  of about 10 km/s at $10^2$~pc, which extends down to scales of about 5~pc. Then, at that scale the turbulent velocity increases to $35-45$~km/s and oscillates around those values until the central region. This represents the turbulent core described by \citet{LatifNature}. Here it begins on the same scale independent of the initial magnetic field used.

The sonic mach number $\mathcal{M}=v/c_s$, where $v$ is the gas velocity magnitude and $c_s$ the sound speed, remains highly supersonic throughout the halo independently of the initial magnetic field, reaching maximum values on the intermediate scales of 60 and in the centre of about 30. This highly supersonic motion is consistent with the convergence of the cold flows from large scales and reflects the turbulent nature of the collapse. 

The turbulent Mach number $\mathcal{M}_{turb}=v_{turb}/c_s$, with $v_{turb}=\sqrt{v_{rms}^2-v_{rot}^2-v_{rad}^2}$ where $v_{rms}$ is the root mean square velocity, $v_{rot}$ is the rotational velocity and $v_{rad}$ is the radial velocity of the gas, also remains high throughout the collapse. At large scales ($r>3$~pc) it reaches values of ~$2.5-10$ and at smaller scales ($r<3$~pc) it reaches values of ~$10-20$. This means that the turbulent pressure dominates over the thermal pressure in providing support against gravitational collapse. The values are rather independent of the initial magnetic field used in our simulations.

The profile showing the vorticity squared displays an almost identical behaviour across for all three initial magnetic field strengths on larger radii ($r>3$~pc). Moving inward, the profile increases gradually toward the center, with a relatively flat behaviour between approximately 0.3~pc and 10 pc. Below $r\sim0.3$~pc, it steepens and the vorticity increases again toward smaller radii. Below $r\sim0.04$~pc, a clear time evolution becomes apparent, where later snapshots (dashed and dotted lines) show systematically higher vorticity than earlier ones (solid lines), reflecting the increase of turbulent motions as the collapse progresses. When comparing the three magnetic field cases at the same evolutionary stage, the profiles remain very similar throughout most of the halo. However, a difference is visible between $0.01$ and $0.3$~pc, where the strongest field case shows slightly higher vorticity. Overall, the vorticity profiles depend more strongly on the time evolution of the collapse rather than the initial magnetic field strength, suggesting that the vorticity generation in this halo is primarily driven by the dynamics of the converging cold streams. This is different from our previous work \citep{diaz}, where the strongest field case showed systematically lower vorticity compared to the weaker field cases. It may suggest that for turbulence at smaller Mach numbers, it is more straightforward for the magnetic field to suppress the vorticity.

The mass accretion rate increases with time, reaching values between 1 and 10~M$_\odot$~yr$^{-1}$ at a peak density  of $10^{-12}$~g~cm$^{-3}$. For the simulation using an initial magnetic field of $B_0=10^{-8}$~G, we see a prominent peak at a radius of $r\sim5\times10^{-2}$~pc, reaching even higher values of about $3\times10^2$~M$_\odot$~yr$^{-1}$, which corresponds to the second clump presented in that simulation.

While most of the physical properties of the halo show a similar behaviour independently of the initial magnetic field strength, the magnetic field does affect the collapse timescales. The onset of the collapse is systematically delayed with increasing field strength. Furthermore, once the collapse is initiated, the stronger field cases take approximately $13$~kyr to evolve from $\rho=3\times10^{-15}$ to $10^{-12}$~g~cm$^{-3}$, compared to only $\sim4$~kyr for the weakest field case. This is consistent with the additional magnetic pressure support provided by the stronger fields, which partially resist the gravitational compression without fundamentally altering the physical properties of the halo.

Figure~\ref{fig:densityproj-002pc} shows the density-weighted projection of the density along the x-axis at a physical scale of $0.02$~pc, for the three initial magnetic field strengths when the simulations reach a peak density of $1\times10^{-12}$~g~cm$^{-3}$. The case with $B_0=10^{-14}$~G shows an elongated central concentration, suggesting the early development of a disc-like structure, while the case with $B_0=10^{-10}$~G shows a more round morphology surrounded by a faint overdense feature on the left side. For $B_0=10^{-8}$~G, the projection shows the most prominent disc-like structure among the three cases, with two spiral arms, similar to the self-gravitating disc identified by \citet{LatifNature} in the purely hydrodynamical simulation of the same halo, though here the morphology is also shaped by the presence of the second clump. The formation of these structures suggests that the magnetic field does not prevent disc formation in this environment.

Overall, this confirms that the magnetic field in this particular scenario has a minimal impact on the morphology of the halo at large scales. The four-stream pattern and irregular shape of the halo remains unchanged if we compare with the simulations presented by \citet{LatifNature} and independently of the initial magnetic field used. Only on scales smaller than 1~pc we can start to notice some differences.

\subsection{Magnetic properties of the halo}\label{magnetic}

Figure~\ref{fig:magnetic64} shows the radial profiles of magnetic field strength, the ratio $B/\rho^{2/3}$ and the magnetic-to-kinetic energy density ratio at the same three evolutionary stages as Figure~\ref{fig:physprop64} for the three simulations. The ratio $B/\rho^{2/3}$ would be expected to be constant under the assumptions of flux freezing and spherically symmetric compression, while it can increase in the presence of strong shocks and non-spherical geometries. 

The magnetic field strength (top panel) increases towards the centre in all three simulations. On larger radii ($r>70$~pc) the field strength  reflects the initial conditions, with a clear separation between the three cases proportional to their different $B_0$ values. Moving inward, all three cases show significant amplification, with the weakest field case ($B_0=10^{-14}$~G) showing the most dramatic increase, particularly between $10 - 70$~pc where its profile rises more steeply than the other two cases. This is unsurprising, as a weaker magnetic field is more strongly affected by the gas flow, provides less damping effects on the gas flow, and can be amplified more strongly by compression or shear especially in the presence of accretion shocks \citep{latif8,Latif2022}. 

At smaller radii, the amplification becomes more gradual and all the three simulations converge to similar central values of $B\sim10^{-2}-10^{-1}$~G by the latest evolutionary stage, despite the initial value of the magnetic field. When comparing the simulations at different times, the central field strength shows a slight increase between the first and second evolutionary stages (solid and dashed lines), of less than an order of magnitude. However, between the second and last evolutionary stages  (dashed and dotted lines), the central field strength shows no further significant evolution, with both lines nearly overlapping in the central region. This behaviour is consistent for all three simulations.
The middle panel shows the ratio $B/\rho^{2/3}$, which helps us to identify magnetic field amplification beyond pure spherical compression. We find that the ratio increases toward the centre in all three simulations, indicating that the magnetic field is being amplified beyond what pure compression alone would predict. However, when comparing different times with fixed initial magnetic field, the ratio decreases in the central region from the first to the later outputs, indicating that while the field increases towards the centre, it grows at a slower rate relative to the density as  the collapse progresses, a result that may both reflect the deviations from spherical symmetry and some backreaction of the magnetic field on the flow. We note that our simulations use a fixed Jeans resolution of 64 cells per Jeans length, which places us at the threshold where small-scale dynamo process may begin to be captured but are likely not fully resolved \citep[see e.g.][]{latif7, grete2019}. 
The bottom panel shows the ratio of the magnetic to kinetic energy density $E_B/E_K$ throughout the halo. In all three simulations and at all evolutionary stages, this ratio remains well below unity, confirming that the magnetic field never becomes dynamically relevant during the collapse. The strongest field cases reach the highest values approaching $E_B/E_K\sim10^{-3}$ in the central regions of the halo, while the weakest field case remains around two orders of magnitude below this. Although the halo is highly turbulent, it is not totally unexpected that the magnetic field does not reach equipartition, as the flow is highly supersonic ($\mathcal{M}\sim30-60$). In such environments, the efficiency of the small-scale dynamo is known to decrease with increasing Mach number in flows dominated by compressible turbulence \citep{brandenburg,Federrath2011,Schober2015}, and our simulations suggest that this effect dominates over any turbulent amplification. This is also consistent with the findings from Figure~\ref{fig:densityproj}, where the physical properties of the halo show no significant dependence on $B_0$. In particular, the low values of $E_B/E_K < 10^{-3}$ found here explain why the magnetic field does not suppress the vorticity, as shown in Figure~\ref{fig:physprop64}, in contrast to \citet{diaz} where the strongest field case reached $E_B/E_K \gtrsim 10^{-1}$, becoming dynamically relevant and reducing the vorticity more efficiently.

We further verify that the magnetic field strength as a function of density converges to comparable central values across the three simulations by the latest evolutionary stage, indicating that the dynamical evolution of the halo, rather than the initial conditions, determines the final field amplitude (see Appendix~\ref{appendix:rhomagnetic}, Figure~\ref{fig:rhomagnetic64}).

Figure~\ref{fig:bproj} shows the density-weighted projection of the magnetic field magnitude along the y-axis at four different spatial scales for the three different initial magnetic fields. At the largest scale ($100$~pc), the magnetic field distribution traces the filamentary structure of the cold streams visible in Figure~\ref{fig:densityproj}, with the strongest field values concentrated in the densest regions. A diagonal structure is visible in all three cases, which does not correspond to the cross-like pattern of the cold streams visible in the density projection. We verified using slices along the same axis that this feature is real and not a projection artifact, though its origin is not entirely clear. It may be related to the interface between two of the converging cold streams, where compression and shear motions could locally amplify the magnetic field.
At intermediate scales ($10$~pc) the morphology of the magnetic field follows the density structure closely, with all three cases showing comparable spatial distribution at this scale.
At $1$~pc, the case with $B_0=10^{-14}$~G shows a somewhat more ordered field distribution compared to the other two cases, which appear more diffuse. This may reflect the chaotic nature of turbulence flow between the three simulations. Additionally, the higher field cases show a closer correspondence between the magnetic field distribution and the underlying density structure, which becomes more apparent at this scale.
At the smallest scale ($0.1$~pc) the simulation with $B_0=10^{-14}$~G and $B_0=10^{-10}$~G show a central concentration with a smooth surrounding gradient. The strongest field case show two distinct concentrations, consistent with the two clumps identified in Figure~\ref{fig:densityproj}. In addition, an enhanced magnetic field is visible between the two clumps, tracing the filament that connects them. The field strengths at this scale are comparable across the three simulations in the centre, consistent with the convergence seen in the radial profiles of Figure~\ref{fig:magnetic64}.

To understand the role of thermal, magnetic and turbulent pressure in determining the mass scale at which the gas can collapse, in addition to the thermal Jeans mass defined in Eq.~\ref{thermaljeans_eq}, we define the magnetic \citep{Mestel1956,Schleicher2010} and turbulent Jeans masses \citep{MacLow2004} as

\begin{equation}
    M_\mathrm{J,mag} = \frac{\pi}{6}\frac{v_\mathrm{A}^3}{G^{3/2}\rho^{1/2}},
\end{equation}
\begin{equation}
    M_\mathrm{J,turb} = \frac{\pi}{6}\frac{v_{turb}^3}{G^{3/2}\rho^{1/2}},
\end{equation}
where $v_\mathrm{A}$ is the Alfv\'en speed.

Figure~\ref{fig:mass64} shows the radial profiles of the thermal Jeans mass (dashed lines), magnetic Jeans mass (solid lines) and turbulent Jeans mass (dotted lines) at the three evolutionary stages.

The thermal Jeans mass decreases toward the centre in all three panels, reflecting the increasing density in the central regions, and is nearly identical across the three different initial magnetic fields at all radii and evolutionary stages.
The magnetic Jeans mass exceeds the thermal Jeans mass only on a limited range, approximately between $0.1$ and $3$~pc just by an order of magnitude in our two simulations with the highest initial magnetic field ($B=10^{-10}$ and $10^{-8}$~G). Outside this range, at larger radii and in the innermost regions, the magnetic Jeans mass falls below or becomes comparable to the thermal value. This behaviour is consistent across all three evolutionary stages. In the simulation with the weakest initial magnetic field ($B_0=10^{-14}$~G), the magnetic and thermal Jeans masses are initially comparable on the $\sim1$~pc scale, though as the collapse progresses, this range extends from $10^{-2}-3$~pc when reaching a peak density of $10^{-12}$~g~cm$^{-3}$. The magnetic Jeans mass, however, never exceeds the thermal Jeans mass. These results indicate that regardless of the initial magnetic field strength, magnetic pressure provides only a minor contribution to the total support against collapse.

The turbulent Jeans mass dominates over both, the thermal and magnetic contributions at all radii, reaching values of $\sim10^{4}$~M$_\odot$ in the central regions, consistent with the estimate by \citet{LatifNature}. This large turbulent Jeans mass reflects the extreme supersonic turbulence driven by the converging cold streams, which prevent the gas from collapsing on small mass scales and explains the formation of massive objects rather than low-mass fragments. The turbulent Jeans mass profiles are also nearly identical across the three initial magnetic field cases, confirming that the initial magnetic field strength does not affect the turbulent dynamics of the halo. 

In addition, the enclosed gas mass profiles provided in Appendix~\ref{appendix:enclosedgasmass} are consistent with the turbulence-dominated dynamics described above (see Figure~\ref{fig:enclosedmass64}), confirming that the global mass distribution is not significantly affected by the initial magnetic field strength.

Overall, these results confirm that gravitational collapse in this halo is determined primarily by turbulent pressure support, with magnetic fields playing a negligible role in setting the characteristic mass scale, in contrast to what has been found in other environments of massive primordial halos where magnetic fields can be amplified strongly enough to significantly suppress fragmentation \citep[e.g.][]{Turk2012,  latif7,  grete2019, diaz}.

%-------------------------------------------------------------
%                                                   CONLUSIONS
%-------------------------------------------------------------

\section{Summary and conclusions}\label{summary}

We have presented cosmological MHD zoom-in simulations of a high-sigma peak that leads to the formation of a dark matter halo with $1.4\times10^{12}$~M$_\odot$ at $z=6$ in a cosmological simulation box of $25$~Mpc~h$^{-1}$. \citet{LatifNature} have previously examined this configuration, showing that strong supersonic turbulence is driven in the intergalactic medium as a result of strong inflows from converging filaments at the halo, which is at the intersection point of these filaments. We have explored the effect of three different initial magnetic fields strengths of $10^{-14}$~G, $10^{-10}$~G and $10^{-8}$~G. The resolution per Jeans length in our simulation was moderate; so we do not expect to resolve significant amplification due to a small-scale dynamo \citep[see e.g.][]{Turk2012, latif7, latif8, grete2019, diaz}, but rather to assess the implications of an initially present magnetic field. Also from an astrophysical consideration, we note that it is more difficult to drive a small-scale dynamo via highly supersonic turbulence \citep[see][]{Federrath2011, Schober2015}. Nonetheless, a comparable Jeans resolution has been shown to be sufficient to capture strong magnetic field amplification effects in more typical atomic cooling halos \citep[e.g.][]{Latif2022}, suggesting that the physical conditions of this halo, rather than the adopted resolution alone, may play a significant role in the magnetic field amplification found here, although a dedicated convergence study would be needed to confirm this directly.

In the simulations provided here, it is nonetheless notable that the magnetic field with the initially lower field strength grows faster as compared to the simulations with stronger initial fields. The latter can be attributed to the fact that a weak field is more susceptible to the flow, allowing stronger amplification through shear, and providing only a small to negligible backreaction onto the flow and basically no damping of the flow itself. It is thus natural to expect some differences in the growth of the magnetic field during the collapse, and while the magnetic fields start out at rather different values, they converge towards a more similar behaviour as a result of the dynamical evolution. Furthermore, the magnetic field affects the collapse timescales, with stronger fields systematically delaying the onset of collapse and slowing the subsequent evolution, as the associated magnetic pressure provides additional support against gravitational collapse.

Turbulent velocity and vorticity do not strongly depend on the initial magnetic field, but rather show variations and fluctuations as a result of the dynamical evolution. Considering the turbulent Mach numbers are in the range of $10-20$, and the overall sonic Mach numbers of the flow in the range of $30-60$, the strongly supersonic flows overall dominate over the possible effects of the magnetic field. This is visible also from an analysis of the characteristic Jeans length, where the largest Jeans mass on all scales is the turbulent Jeans mass. Thermal and magnetic Jeans mass are always subdominant compared to the turbulent Jeans mass. On scales of $\sim10$~pc down to $10^{-2}$~pc, the magnetic Jeans mass can be comparable to the thermal Jeans mass, particularly in the simulations starting from a stronger initial field. In the simulation with the weakest initial field of $10^{-14}$~G, they are initially only comparable on the $\sim1$~pc scale, though when the configuration reaches a higher density peak of $10^{-12}$~g~cm$^{-3}$, they become comparable over a dynamic range from $10^{-2}-3$~pc. This confirms that, overall in this highly supersonic regime, while the magnetic field can be amplified to a certain extent, the turbulence remains dynamically more important than the magnetic contribution.

The results are thus considerably different compared to simulations of more typical atomic cooling halos \citep[e.g.][]{latif7, grete2019, diaz}, where supersonic flows also exist due to the virialisation shocks, but on a much more moderate level. Particularly, the magnetic Jeans mass was shown to clearly dominate on scales of $0.1-3$~pc in such standard halos \citep{Latif2022, Latif2023}, while here the turbulent Jeans mass is always by far the largest Jeans mass and only occasionally the magnetic Jeans mass becomes comparable to the thermal one, while both remain subdominant compared to the turbulent Jeans mass. 

This is further reflected in the small-scale morphology of the halo. The presence of disc-like structures at sub-pc scales is consistent with \citet{LatifNature}, who identified a similar structure in their purely hydrodynamical simulation of the same halo. As noted by \citet{LatifNature}, the extreme turbulence in this environment prevents the formation of large coherent accretion discs typically seen in atomic cooling halos, though smaller disc-like structures can still develop within the dense clumps. In atomic cooling halos, magnetic fields have been shown to stabilize accretion discs by suppressing fragmentation through magnetic pressure and angular momentum transport \citep{Latif2022,Latif2023}. In the present case, however, the turbulence dominates over magnetic fields throughout the collapse, and the disc-like structures that form in our simulations are therefore primarily a consequence of the turbulent and gravitational dynamics driven by the converging cold streams, independently of the initial magnetic field strength.

The fact that we find evidence for a self-gravitating accretion disc in this extreme halo also gives rise to further possibilities for magnetic field amplification via magnetic dynamos within the disc \citep[e.g.][]{brandenburg, Latif2016M}. Such magnetized discs could efficiently drive jets and provide a feedback mechanism within the galaxy, while the magnetic fields could also contribute to the transport of angular momentum \citep{Begelman2023}. In other contexts, it has also been suggested that the magneto-rotational instability could affect magnetized discs, though the latter is more likely if the magnetic fields are relatively weak \citep{Silk2006}. Particularly in the case of a self-gravitating disc, it is then important to consider the interaction of the magneto-rotational instability with the self-gravitating instability \citep{Fromang2004}.

We also note that the strongest-field simulation formed a binary structure, unlike the other two cases, which may indicate that a fraction of such configurations produces binaries, though our low-number statistics cannot determine whether this behaviour has a physical origin related to the magnetic field or is instead a consequence of stochastic variations in the collapse. Therefore, a larger simulation sample of these high-sigma peak environments will be needed to clarify this.

The result here is not necessarily unexpected but shows that magnetic field amplification and its dynamical effects can depend on the specific physical conditions of the halo considered, such as the extreme turbulence found here. Particularly in the presence of strong cosmic inflows through filaments, our simulations strongly suggest that these dynamical influences are regulating the flow properties. While the small-scale dynamo has been shown to amplify magnetic fields close to saturation in more typical atomic cooling halos \citep{grete2019,Latif2022,diaz}, the amplification found here is not sufficient to compete with the dominant turbulence of this particular halo. Nonetheless, it is still conceivable that, particularly during the formation of a supermassive star or the collapse of a larger part of the gas towards a supermassive black hole, the magnetic field still could have a stabilizing role to reduce the amount of fragmentation. Once that a supermassive star has formed, magnetic fields could also regulate its evolution through effects such as magnetic braking \citep{Haemmerle2019}. For the rare sigma peaks considered here, a relevant question concerns the evolution after the formation of a supermassive black hole, where potentially winds and jets from the accretion disc could start providing feedback in the presence of a magnetic field. It seems likely that even in this case, efficient accretion would be possible in such a scenario, given the strongly convergent inflows on a cosmic scale. Nonetheless, it will be important to further examine the evolution of these rare objects and what may happen under such extreme conditions.

\begin{acknowledgements}
 We thank Philipp Grete for useful discussions. VBD acknowledges financial support from ANID (ANID-PFCHA/DOCTORADO DAAD-BECAS CHILE/62200025) as well as financial support from DAAD (DAAD/Becas Chile funding program ID 57559515). The authors gratefully acknowledge the computing time granted by the Resource Allocation Board and provided on the supercomputer Lise and Emmy at NHR@ZIB and NHR@Göttingen as part of the NHR infrastructure. The calculations for this research were conducted with computing resources under the project hhp00057.  DRGS gratefully acknowledges support by the ANID BASAL projects ACE210002 and FB210003 and via the Alexander von Humboldt - Foundation, Bonn, Germany. MAL thanks the UAEU for funding via UPAR grants No. 31S390 and 12S111. RB acknowledges support by the Deutsche Forschungsgemeinschaft (DFG, German Research Foundation) under Germany’s Excellence Strategy – EXC 2121 „Quantum Universe“ – 390833306. The visualization and analysis of this research was done thanks to the YT project, an open-source, community-developed python package for astrophysical data \citep{YTproject}. Parts of the results in this work make use of the colormaps in the CMasher package \citep{cmasher}.
\end{acknowledgements}


\bibliographystyle{aa} % style aa.bst
\bibliography{aanda} % your references Yourfile.bib

%-------------------------------------------------------------
%               Appendices have to be placed at the end, after
%                                        \end{thebibliography}
%-------------------------------------------------------------

\begin{appendix}
\onecolumn
\section{Density profiles of magnetic field strength}
\label{appendix:rhomagnetic}

\begin{figure}[H]
   \centering
   \includegraphics[width=0.55\hsize]{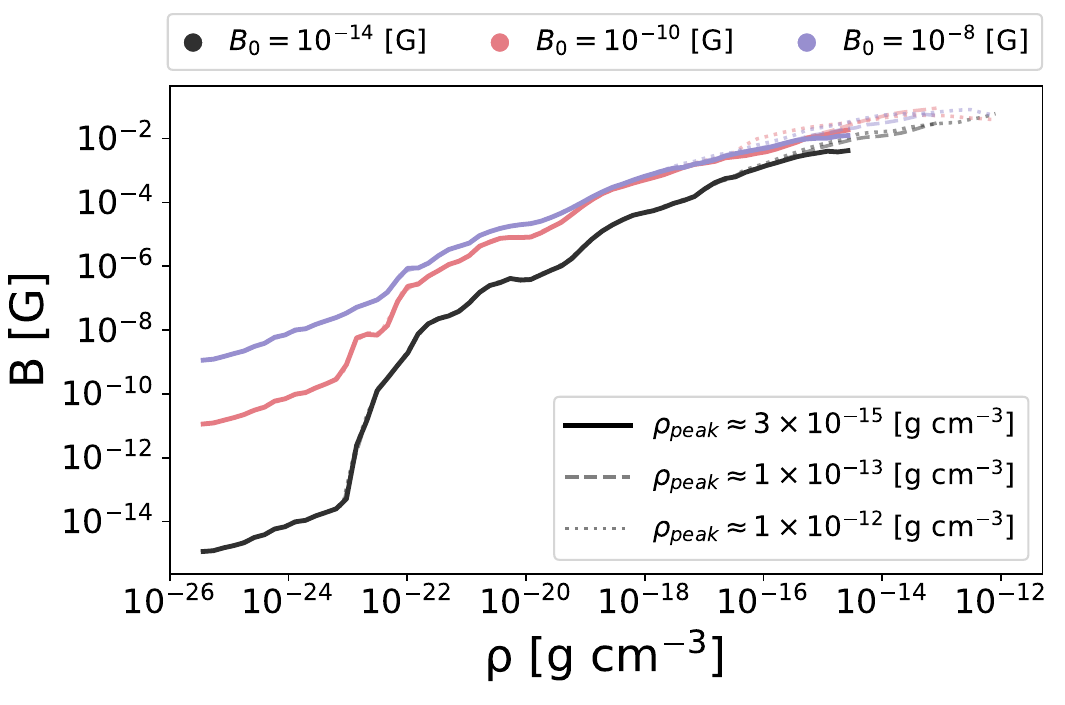}
   \caption{Magnetic field strength plotted against density when reaching three different peak densities. Black lines are for runs with $B_0=10^{-14}$~[G] (proper), pink lines for runs with $B_0=10^{-10}$~[G] (proper), and purple lines for runs with $B_0=10^{-8}$~[G] (proper). Solid lines correspond to $\rho_{peak}=3\times10^{-15}$~$\mathrm{[g\,cm^{-3}]}$, dashed lines correspond to $\rho_{peak}=1\times10^{-13}$~$\mathrm{[g\,cm^{-3}]}$ and dotted lines correspond to $\rho_{peak}=1\times10^{-12}$~$\mathrm{[g\,cm^{-3}]}$. Despite the initial magnetic field strengths, all three simulations converge to similar values of $B\sim10^{-2}$~G at the highest densities by the latest evolutionary stage. The slope of the profiles is steeper than the $B\propto\rho^{2/3}$ expectation from pure compression, indicating amplification beyond flux freezing.
   }
    \label{fig:rhomagnetic64}
    \end{figure}

Figure~\ref{fig:rhomagnetic64} shows the magnetic field strength as a function of density for the three simulations at the three evolutionary stages. At low densities, the three cases are clearly separated, reflecting their initial field strengths. Moving towards higher densities, the profiles steepen beyond $B\propto\rho^{2/3}$, with the weakest field case showing the most dramatic rise, consistent with the stronger amplification observed at larger radii in Figure~\ref{fig:magnetic64}. By the latest evolutionary stage (dotted lines), all three cases converge to a field strength of $B\sim10^{-2}$~G despite the different initial values, further confirming that the final field amplitude is set by the dynamical evolution of the halo rather than the initial conditions.
%-------------------------------------- Two column figure (place early!)
%-----------------------------------------------------------------

\newpage
\section{Enclosed gas mass profiles}
\label{appendix:enclosedgasmass}

\begin{figure}[H]
   \centering
   \includegraphics[width=0.55\hsize]{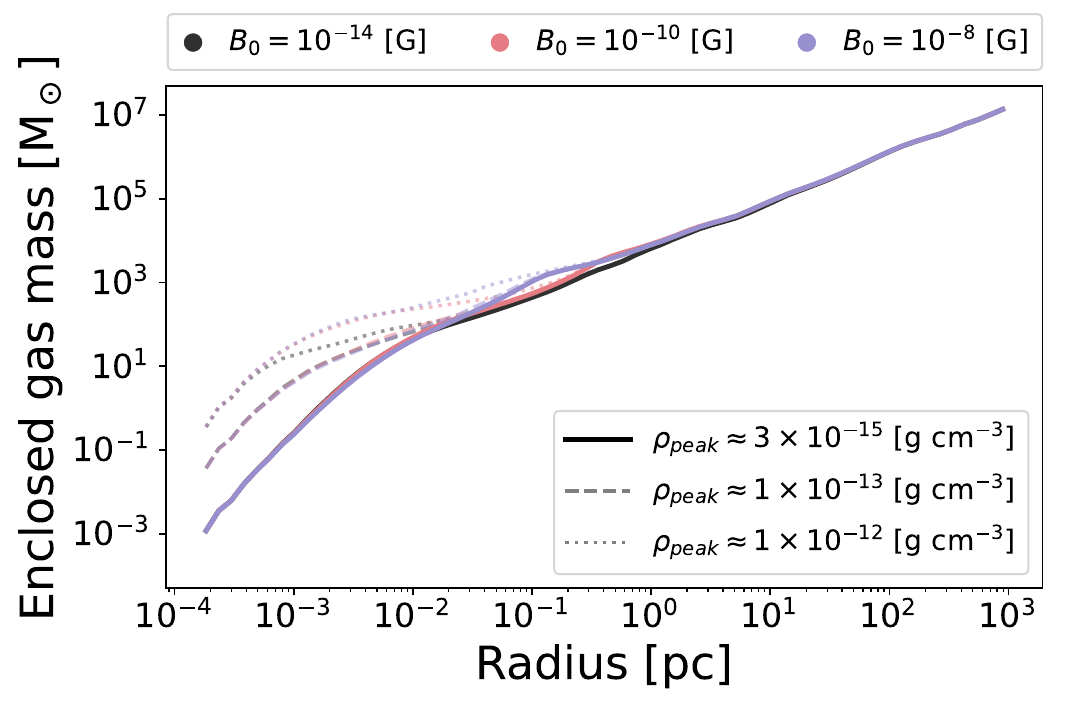}
   \caption{Mass-weighted spherically binned radial profiles of enclosed gas mass when reaching three different peak densities. Black lines are for runs with $B_0=10^{-14}$~[G] (proper), pink lines for runs with $B_0=10^{-10}$~[G] (proper), and purple lines for runs with $B_0=10^{-8}$~[G] (proper). Solid lines correspond to $\rho_{peak}=3\times10^{-15}$~$\mathrm{[g\,cm^{-3}]}$, dashed lines correspond to $\rho_{peak}=1\times10^{-13}$~$\mathrm{[g\,cm^{-3}]}$ and dotted lines correspond to $\rho_{peak}=1\times10^{-12}$~$\mathrm{[g\,cm^{-3}]}$. The enclosed gas mass of the halo is nearly identical across all three initial magnetic field strengths at large scales, confirming that the global mass distribution is unaffected by the magnetic field. At smaller scales, the enclosed mass increases with time as the collapse progresses, with all three cases reaching similar values at each evolutionary stage, consistent with the turbulence-dominated dynamics of the halo.
   }
    \label{fig:enclosedmass64}
    \end{figure}

Figure~\ref{fig:enclosedmass64} shows the mass-weighted spherically binned radial profiles of enclosed mass at three evolutionary stages. At the radii where the turbulent Jeans mass reaches $\sim10^{4}$~M$_\odot$, the enclosed gas mass of the halo increases during the collapse. However, by our latest evolutionary stage, it remains considerably smaller than this threshold. This confirms that the gas at those scales is unable to fragment into low-mass objects at that stage of the collapse, independently of the initial magnetic field strength, as the characteristic mass scale for collapse at these radii is set by turbulent pressure rather than by magnetic or thermal support. At larger radii ($r>2\times10^{-1}$~pc), the enclosed gas mass increases sharply and is nearly identical across all three initial magnetic field strengths and across all evolutionary stages, showing that the large-scale mass distribution is not affected by the initial magnetic field strength.
%-------------------------------------- Two column figure (place early!)
%-----------------------------------------------------------------
\end{appendix}

\end{document}